\documentclass[aps,prl,reprint,amsmath,amssymb]{revtex4-1}

\usepackage[english]{babel}

\usepackage[cp1251]{inputenc}
\usepackage[mathlines]{lineno}
\usepackage{amsmath,amssymb,graphicx,color}

\usepackage{amssymb,amsmath,latexsym}
\usepackage{setspace}
\usepackage{braket}
\usepackage{graphicx}
\graphicspath{{./Figs/}}
\usepackage{dcolumn}
\usepackage{bm}
\usepackage{hyperref}
\usepackage{amsfonts}
\usepackage{amsmath,amssymb}
\usepackage{amsmath,amsthm}
\usepackage{amssymb}
\usepackage{graphicx,color}
\usepackage{subfigure}
\usepackage{epstopdf}
\usepackage{xcolor}
\usepackage{color}
\usepackage{natbib}
\usepackage[strict]{changepage}

\usepackage{epsfig}
\usepackage{fancybox}
\usepackage{listings}
\usepackage{url}
\usepackage{verbatim}
\newcommand{\eval}[2][\right]{\relax \ifx#1\right\relax \left.\fi#2#1\rvert}

 \def\be{\begin{equation}}
 \def\ee{\end{equation}}
 \def\bes{\begin{eqnarray}}
 \def\ees{\end{eqnarray}}

 \def\2{\frac{1}{2}}
 \def\4{\frac{1}{4}}

\begin{document}
\title{Optically assisted trapping with high-permittivity dielectric rings: Towards optical aerosol filtration}
\author{Rasoul Alaee} 
\affiliation{Institute of Theoretical Solid State Physics, Karlsruhe Institute of Technology, Karlsruhe 76131, Germany}
\affiliation{Max Planck Institute for the Science of Light, Erlangen 91058, Germany}
\author{Muamer Kadic} 
\affiliation{Institute of Applied Physics, Karlsruhe Institute of Technology, Karlsruhe 76128, Germany}
\affiliation{Institut FEMTO-ST, Universit\'e  de Bourgogne Franche-Comt\'e, CNRS, 25044 Besancon Cedex, France}
\author{Carsten Rockstuhl}
\affiliation{Institute of Theoretical Solid State Physics, Karlsruhe Institute of Technology, Karlsruhe 76131, Germany}
\affiliation{Institute of Nanotechnology, Karlsruhe Institute of Technology, Karlsruhe 76021, Germany}
\author{Ali Passian} 
\affiliation{Computational Sciences and Engineering Division, Oak Ridge National Laboratory, Oak Ridge, TN 37831-6123, USA}
\affiliation{Department of Physics and Astronomy, The University of Tennessee, Knoxville, TN 37996-1200, USA}

\begin{abstract}
Controlling the transport, trapping, and filtering of nanoparticles is important for many applications. By virtue of their weak response to gravity and their thermal motion, various physical mechanisms can be exploited for such operations on nanoparticles. However, the manipulation based on optical forces is potentially most appealing since it constitutes a highly deterministic approach. Plasmonic nanostructures have been suggested for this purpose, but they possess the disadvantages of locally generating heat and trapping the nanoparticles directly on surface. Here, we propose the use of dielectric rings made of high permittivity materials for trapping nanoparticles. Thanks to their ability to strongly localize the field in space, nanoparticles can be trapped without contact. We use a semi-analytical method to study the ability of these rings to trap nanoparticles. Results are supported by full-wave simulations. Application of the trapping concept to nanoparticle filtration is suggested.
\end{abstract}

\maketitle
The study of single isolated nanoparticles (NPs) and their properties
 is important for understanding how they may mutually agglomerate as well as for how they interact with other types of matter. With the rising commercial use and large-scale production of engineered NPs, methods to capture and filter NPs are needed for environmental and health-care aspects but also to assure a sustainable nanotechnology. Disciplines such as analytical chemistry~\cite{wei} and nanotoxicology~\cite{ai} are concerned with methods to characterize nanomaterials with respect to their impact upon unintended exposure or, more generally, nanotoxicity. To perform research in these fields, access to a large number of NPs and agglomerates in a specific experimental surrounding is required. Therefore, these disciplines would benefit from basic physical processes that are capable to spatially and temporally control the motion of NPs. Going one step further, these processes may also be used in respirator function for protection in pandemic of respiratory diseases (e.g, SARS, bird and swine flu), where inhalation of NPs, which can be easily aerosolized, is of great concern. Trapping and controlling the motion of particles by means of optical forces is one answer to these needs.
\begin{figure}[h!]
    \centering
           \includegraphics[width=0.75\columnwidth]{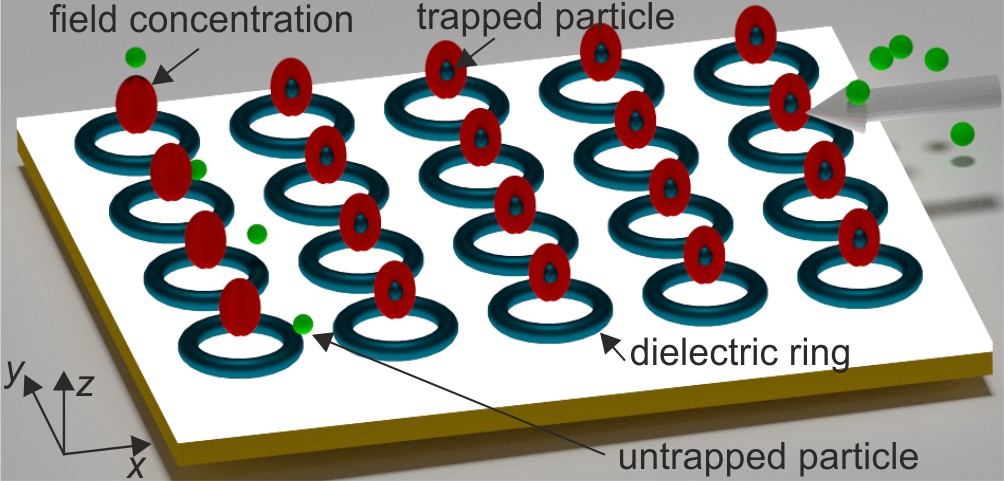}
           \caption{Schematic configuration of the suggested NP trapping concept. Particles (blue) with a dipole moment will be trapped by the force field (red) generated upon illuminating a dielectric ring (blue) by an incident field propagating along the $z-$ direction while other particles (green) remain untrapped.}
    \label{setup}
\end{figure}
The demonstration of optical trapping of dielectric particles in liquids by Ashkin {\it et al.} using a single laser beam~\cite{ashkin-1970} and the extension of the concept to Rayleigh particles and atoms led to many interesting force-gradient based trapping venues. The development of optical trapping~\cite{marago} of small particles led to the development of optical tweezers~\cite{ashkin-1986} and studies of microrheological properties of aerosol particles~\cite{power}, optical binding of trapped particles~\cite{bowman}, and DNA and cell studies~\cite{Bustamante:2008}. Non-contact force gradient approaches based on optical and dielectrophoresis \cite{Pethig2010} for manipulating and controlling nano- or micro-objects are specially useful when contact assisted control such as micromanipulators or atomic force microscopy may be less effective due to electrostatic sticking, contact damage, capillarity, etc.. For example, one may attempt deposition under a contactless configuration.
\begin{figure}[h!]
    \centering
           \includegraphics[width=\columnwidth]{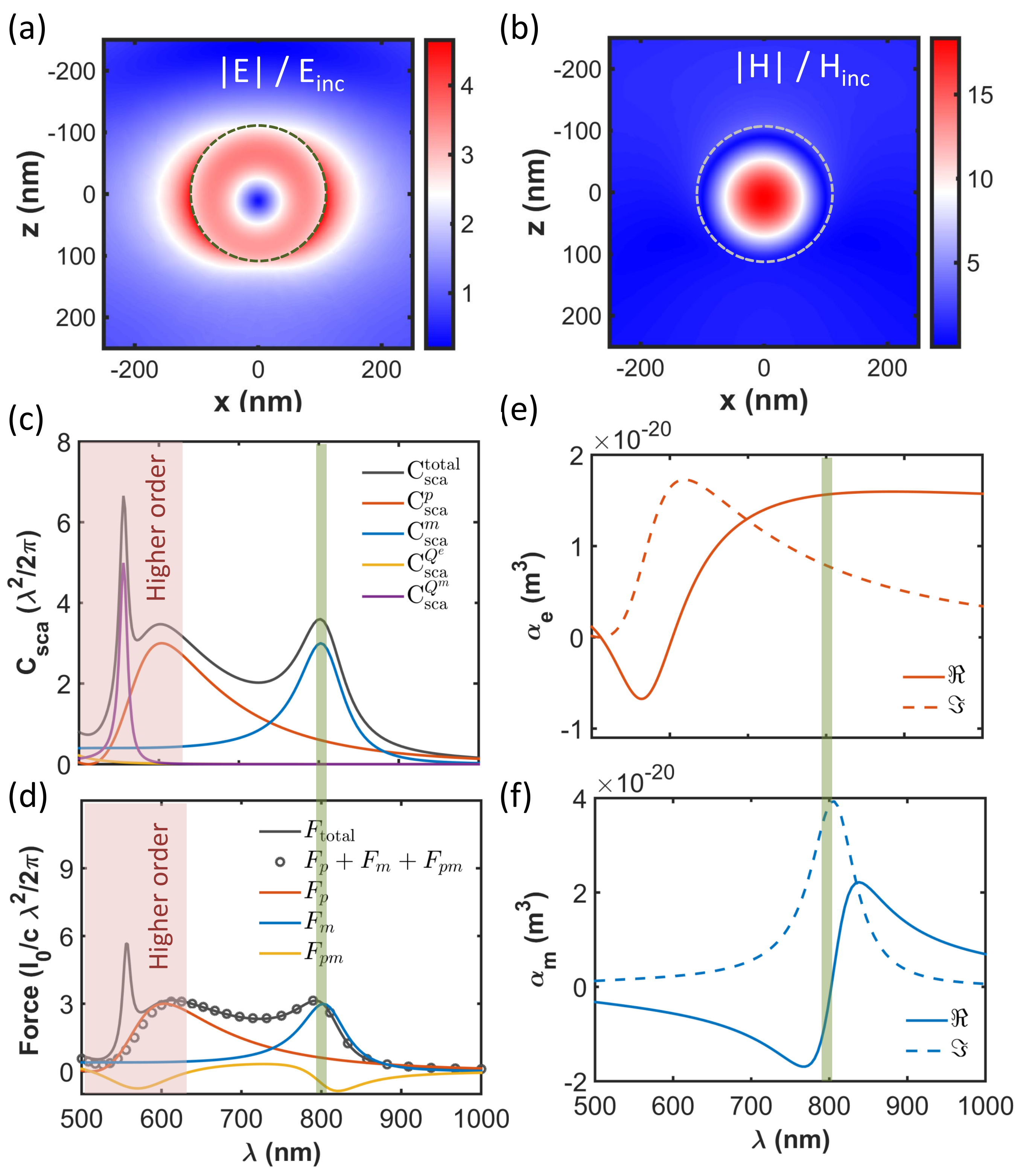}
           \caption{\textit{Dielectric sphere}: Scattering properties of a dielectric NP with a relative permittivity of $\epsilon_r= 12.25$ and radius of 110~nm. (a) and (b) Normalized electric and magnetic amplitude distributions at resonance (\emph{i.e.}, 800~nm)  in the $xz$-plane, respectively. The dielectric sphere is illuminated by a linearly x-polarized plane wave propagating along the z-axis. (c) Total scattering cross section $C_{\mathrm{sca}}^{\mathrm{total}}$
and contributions from different multipole moments as a function of
the wavelength. Contributions are only shown from the electric dipole moment $C_{\mathrm{sca}}^{p}$, magnetic dipole moment $C_{\mathrm{sca}}^{m}$, electric
quadrupole moment $C_{\mathrm{sca}}^{Q^{e}}$, and magnetic
quadrupole moment $C_{\mathrm{sca}}^{Q^{m}}$. (d) Normalized
total optical force $F_{\mathrm{total}}$ and contribution of different
multipoles; electric $F_{p}$, magnetic $F_{m}$, interference term
$F_{pm}$ exerted on the particle. (e) and (f) Electric $\alpha_{\mathrm{e}}$ and
magnetic $\alpha_{\mathrm{m}}$ polarizabilities of the spheres.}
    \label{fig:Sphere}
\end{figure}
Here, by proposing the contactless purely dielectric trapping configuration depicted in Fig.~\ref{setup}, we show that a ring made from a high permittivity dielectric material creates a spatial region with sufficiently large field gradient where NPs can be stably trapped.
Both, a single trap and an array of traps may be envisioned as an active contactless electromagnetically controlled filter. Our work exploits dielectrics as the material from which the ring is made. This is in contrast to metallic rings studied in literature \cite{Aizpurua2003,Juan:2011,Lehr:2014,Alaee:2015}. However, scattering nanostructures from high permittivity dielectric materials were recently identified as a suitable platform to replace plasmonic structures in selected applications~\cite{Krasnok2012,Staude2013,Odit:2016,Jahani:2016}. Thanks to the extraordinary control of the multipolar composition of the scattered field combined with the vanishing absorption, dielectric structures are appealing in many instances. Here, we explore the possibility of using such high permittivity scatterers to trap NPs.
Here, air is assumed as the host ambient medium and, as depicted in Fig.~\ref{setup},  one may introduce a substrate, which other than lowering the trapping frequencies slightly will have no other qualitative effects.
In ambient air, the mean free path of the air molecules is $\lambda_{\mathrm{mfp}}\approx 65$~nm. Therefore, for particles with a diameter $r_{\mathrm{p}}\propto \lambda_{\mathrm{mfp}}$, the molecule-particle surface collisions are not small compared to molecule-molecule collisions~\cite{Passian:2002,Passian:2003}. Thus, without elaborating, it is reasonable to assume that a particle of mass $m$ at a position ${\bf u}(t)$ will obey the Langevin equation $m\ddot{{\bf u}}(t)=\nabla (mg {\bf u} (t)+ \int_{{\bf u}_0}^ {\bf u} {\bf F} d{\bf u})- \gamma \dot{{\bf u}}(t) + {\bf \Xi}$, where $g$ and $\gamma$ are the constants of gravity and dissipation, and the random force $\mathbf{\Xi}$ can be obtained from the fluctuation-dissipation relation. Our aim is to investigate the optical force $\mathbf{F}$ as a means to trap the NP.
For this purpose, we first introduce the numerical and theoretical approaches to compute the optical force based on Maxwell's stress tensor and the induced multipole moments~\cite{Nieto-Vesperinas2010,Rahimzadegan2016,FernandezCorbaton:15}. We then compute the exerted optical force on an isolated dielectric NP illuminated by a plane wave. Afterwards, we study the interaction of light with an isolated ring and show the presence of a field concentration (hot spot). Having characterized the field in close proximity to the ring, we place a spherical dielectric NP at various positions and calculate the force exerted on it. This is obtained semi-analytically within dipole approximation while neglecting the scattered field generated by the NP (\emph{i.e.}, Eqs.~\ref{eq:ForceTS}-\ref{eq:ForceTE}). Moreover, the force is also calculated from first principles while considering the self-consistent field in the nanoparticle-ring configuration and by integrating Maxwell's stress tensor (\emph{i.e.}, Eq.~\ref{eq:ST_def}).
Both approaches are shown to be in good agreement. Finally, we obtain the trajectory of the NP by considering various initial conditions and show that for the induced field, trapping is feasible.

We begin by noting  that for a host medium characterized by its permittivity $\epsilon$ and permeability $\mu$, the time averaged optical force exerted on an arbitrarily shaped NP is given by
$\mathbf{F}  =  \oint_{S}\mathbf{T}\cdot\mathbf{n}da$, where $S$ is a surface enclosing the particle,  $\mathbf{n}$ is the normal to
$S$, and $\mathbf{T}$ is the time-averaged Maxwell's stress tensor:
\begin{eqnarray}
\mathbf{T} & = & \frac{1}{2}\mathrm{Re}\left[\epsilon\mathbf{E}\mathbf{E}^{*}+\mu\mathbf{H}\mathbf{H}^{*}-\frac{1}{2}\left(\epsilon\mathbf{E}\cdot\mathbf{E}^{*}+\mu\mathbf{H}\cdot\mathbf{H}^{*}\right)\mathbf{I}\right].
\label{eq:ST_def}
\end{eqnarray}
with  $\mathbf{E}$ and $\mathbf{H}$ being the complex total electric and magnetic fields, and $\mathbf{I}$ being the identity matrix~\cite{Barton1989,almaas1995,Jackson1999}.
Within the dipole approximation however, the time averaged force induced by an arbitrary incident wave may be written as:
\begin{eqnarray}
\mathbf{F} & = & \mathbf{F}_{p}+\mathbf{F}_{m}+\mathbf{F}_{pm},\label{eq:ForceTS}\\
\mathbf{F}_{p} & = & \frac{1}{2}\mathrm{Re}\left[\left(\nabla\mathbf{E}_{\mathrm{inc}}^{\mathrm{*}}\right)\cdot\mathbf{p}\right],\\
\mathbf{F}_{m} & = & \frac{1}{2}\mathrm{Re}\left[\left(\nabla\mathbf{B}_{\mathrm{inc}}^{\mathrm{*}}\right)\cdot\mathbf{m}\right],\\
\mathbf{F}_{pm} & = & -\frac{Z_{0}k^{4}}{12\pi}\mathrm{Re}\left[\left(\mathbf{p}\times\mathbf{m}^{*}\right)\right],\label{eq:ForceTE}
\end{eqnarray}
where $\mathbf{F}_p$, $\mathbf{F}_m$, and $\mathbf{F}_{pm}$ are forces resulting from the induced electric  dipole moment $\mathrm{\mathbf{p}}$, magnetic
 dipole moment $\mathrm{\mathbf{m}}$, and their mutual interference~\cite{chen2011, Nieto-Vesperinas2010,Rahimzadegan2016}.
Denoting the electric and magnetic polarizabilities with $\alpha_e$ and $\alpha_m$, respectively, the induced moments are
$\mathbf{p}=\epsilon_{0}\alpha_{\mathrm{e}}\mathbf{E}_{\mathrm{inc}},$ and $\mathbf{m}=\alpha_{\mathrm{m}}\mathbf{H}_{\mathrm{inc}}$. Both $\alpha_e$ and $\alpha_m$ are assumed scalar quantities due to the isotropicity of the spherical NPs considered.

\begin{figure}[h!]
    \centering
           \includegraphics[width=0.9\columnwidth]{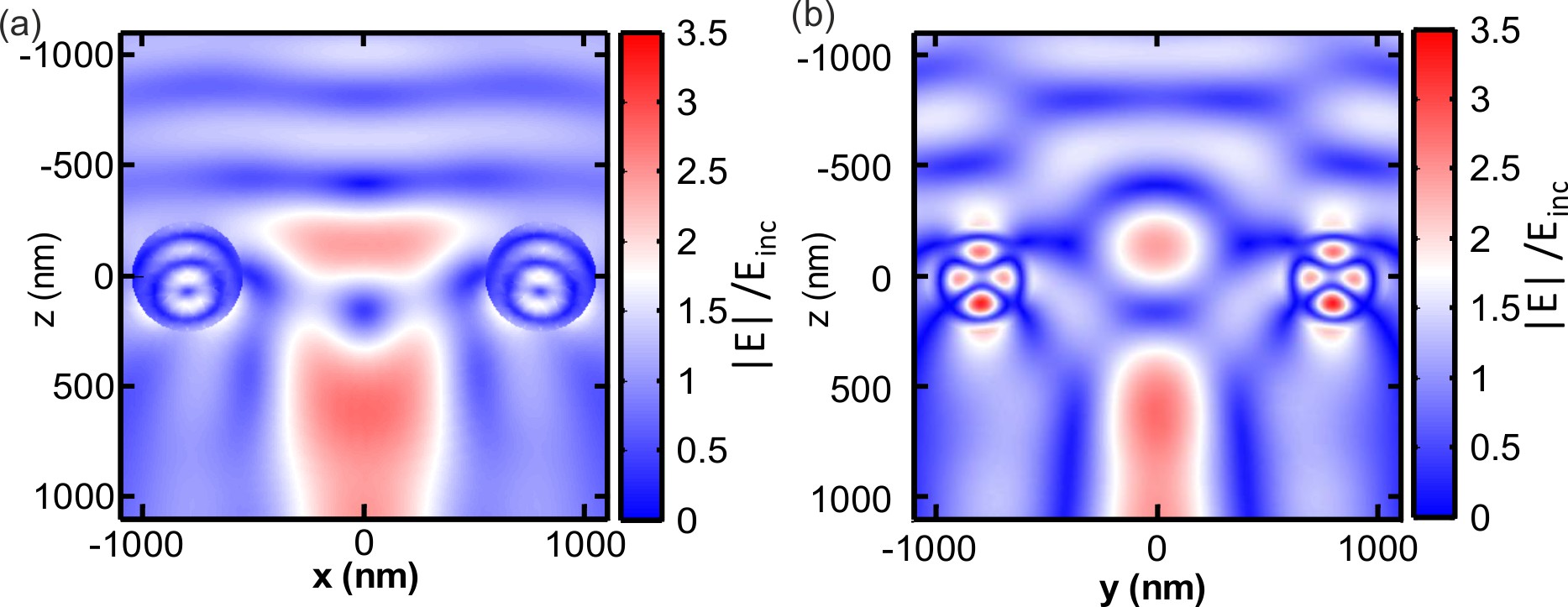}
           \caption[]{\textit{High permittivity dielectric ring}: Illustration of the field distribution upon illumination with a linearly $x$-polarized plane wave that propagates in the $z$ direction. (a) and (b) Field distributions in the $xz-$ and $yz$-plane, respectively. The relative permittivity of ring is $\epsilon_r= 12.25$. The major and minor radius of the ring are 800~nm, 250~nm, respectively.}
     \label{fig:Torus}
\end{figure}
Noting that a high permittivity dielectric NP yields both an electric as well as a magnetic dipolar response~\cite{Kuznetsov2012,Krasnok2012,Staude2013,Odit:2016}, we employ the Mie theory~\cite{Bohren2008} and compute the total scattering cross section and the contribution of each multipole moment, as shown in Fig.~\ref{fig:Sphere}. The higher order moments, \emph{i.e.}, electric quadrupole $C_{\mathrm{sca}}^{Q^e}$ and magnetic quadrupole $C_{\mathrm{sca}}^{Q^m}$ moments are considerable only below a wavelength of 650~nm, as indicated by the red shaded areas in Fig.~\ref{fig:Sphere}.
As also indicated in Fig.~\ref{fig:Sphere}, by fixing the wavelength at 800~nm, we here choose to work within a spectral range where the NP can only support electric and magnetic dipole moments [Fig.~\ref{fig:Sphere} (a)-(c)]. Therefore, the force is sufficiently described by the dipolar expressions given by Eq.~\ref{eq:ForceTS}. Figure~\ref{fig:Sphere}(e)-(f) show the induced electric and magnetic polarizabilities, while  Fig.~\ref{fig:Sphere}(d) displays the total optical force $F_{\mathrm{total}}$ computed from the time-averaged Maxwell stress tensor, \emph{i.e.}, Eq.~\ref{eq:ST_def} for a plane wave illumination. Furthermore, using Eqs.~\ref{eq:ForceTS}-\ref{eq:ForceTE}, the corresponding total optical force and the contribution of electric, magnetic, and interference terms are displayed in Fig.~\ref{fig:Sphere} (d). Note that the force and scattering cross sections are normalized because there is a universal limit on the scattering cross section~ \cite{Ruan2010,Ruan2011} $C_{\mathrm{max}}=\frac{3\lambda^{2}}{2\pi}$, similarly for
the optical force $F_{\mathrm{max}}=\frac{I_{0}}{c}\frac{3\lambda^{2}}{2\pi}$  of a dipolar scatter~\cite{Rahimzadegan2016}, where $I_0$ is the intensity of the impinging plane wave.
\begin{figure}
    \centering
           \includegraphics[width=0.95\columnwidth]{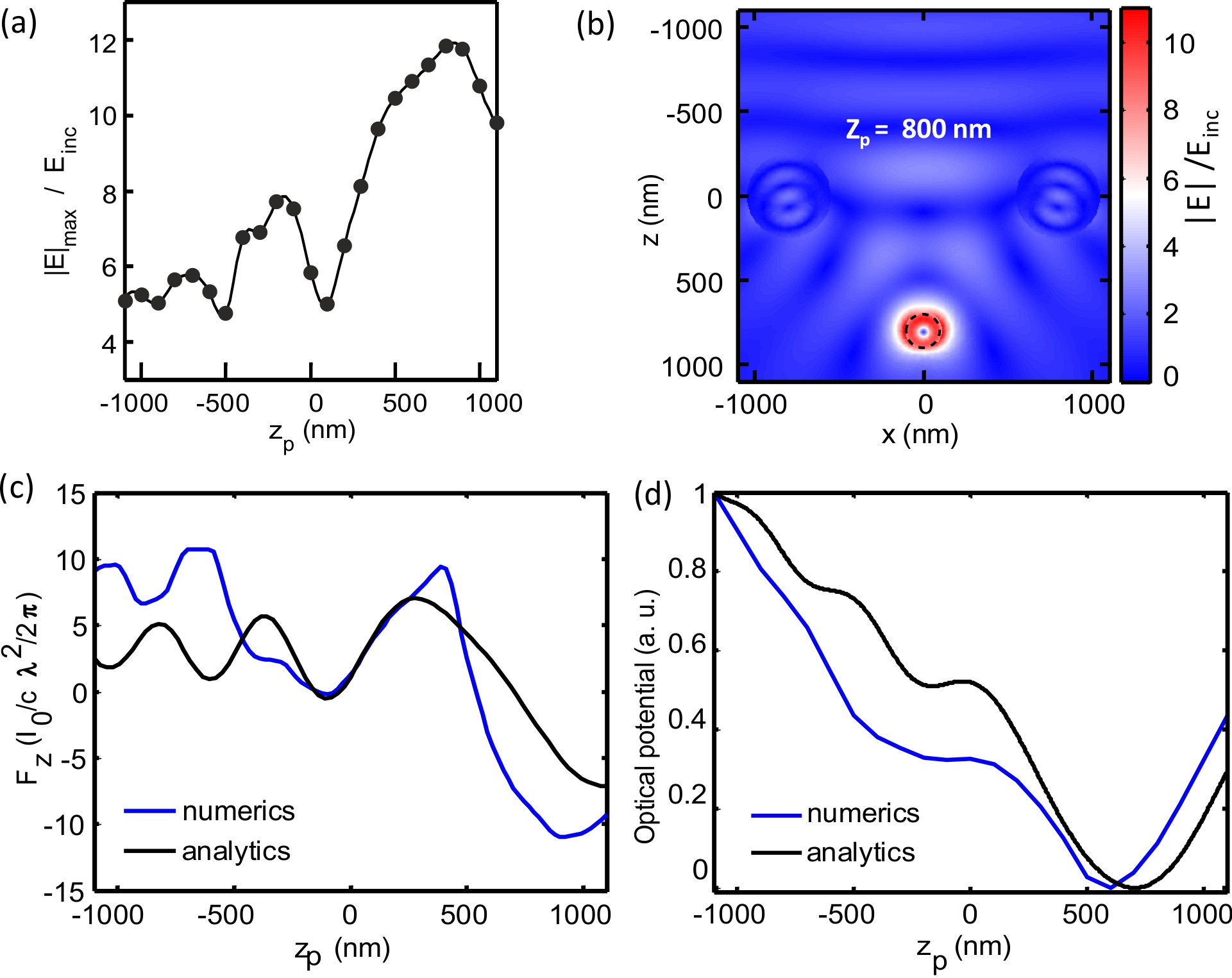}
           \caption[]{Nanoparticle-ring configuration upon illumination with a linearly $x$-polarized plane wave that propagates in the $z$ direction. (a) Variation of the maximum electric field amplitude (calculated in the xz-plane) when the particle is displaced along the $z$ axis for $x=y=0$  (b) Amplitude distribution of the electric field in the xz-plane for $z_p=800$~nm of the NP.
           Position dependent exerted force on the dielectric NP.
                      (c) $z$-component of the optical force as
           a function of particle position $z_{\mathrm{p}}$. (d) The corresponding
           optical potential, which is normalized to its maximum. The optical potential is calculated by using the relation between the optical force and potential, \emph{i.e.},  $F_z=-\frac{dU}{dz}$, where U is the optical potential.}
     \label{fig:TorusWithSphere}
\end{figure}
The response of an isolated ring to an $x$-polarized field is first obtained computationally using the finite elements method~\cite{comsol}. The ability of the dielectric ring to concentrate the energy into a finite spatial region that can be considered as a focusing effect is shown in Fig.~\ref{fig:Torus} (a)-(b). The emergence of a hot spot can be observed about 700 to 800~nm below the ring. If we now assume the initial NP position $u|_{t=0}$ to be  along the $z$-axis, we observe that the maximum field (in the $xz$-plane) is formed when the particle is \textit{approximately} within the hot spot region (\emph{i.e.}, $z_\mathrm{p}\approx$ 800 nm), as shown in Fig.~\ref{fig:TorusWithSphere} (a). For the chosen materials and particles, it is reasonable to assume that the scattered field generated by the  NP is negligible when compared to the field generated by the ring itself. Thus, we may obtain the force on the spherical NP semi-analytically by using Eqs.~\ref{eq:ForceTS}-\ref{eq:ForceTE}, where the electric and magnetic fields are only due to the ring (shown in Fig.~\ref{fig:Torus}).
However, this approximation is intended to simplify the computation of the force and therefore warrants a validation. Therefore, we also calculate the force  by using the self-consistent fields of the nanoparticle-ring configuration (shown in Fig.~\ref{fig:TorusWithSphere}) via integrating Maxwell's stress tensor \emph{i.e.}, Eq.~\ref{eq:ST_def}. A comparison between the exerted force on the NP as well as the optical trapping potential calculated by the two approaches is shown in Fig.~\ref{fig:TorusWithSphere} (c) and (d). Except for minor shifts, all features are qualitatively and quantitatively present. Having established that the semi-analytical method is sufficiently accurate, we will now proceed to obtain the trajectory of the NP.
Here, the utility of the analytical results obtained can be appreciated when noting that  solving the coupled Navier-Maxwell equations, using frequency (Maxwell) and time (Navier) domains, require significant computational resource.

\begin{figure}
    \centering
           \includegraphics[width=0.6\columnwidth]{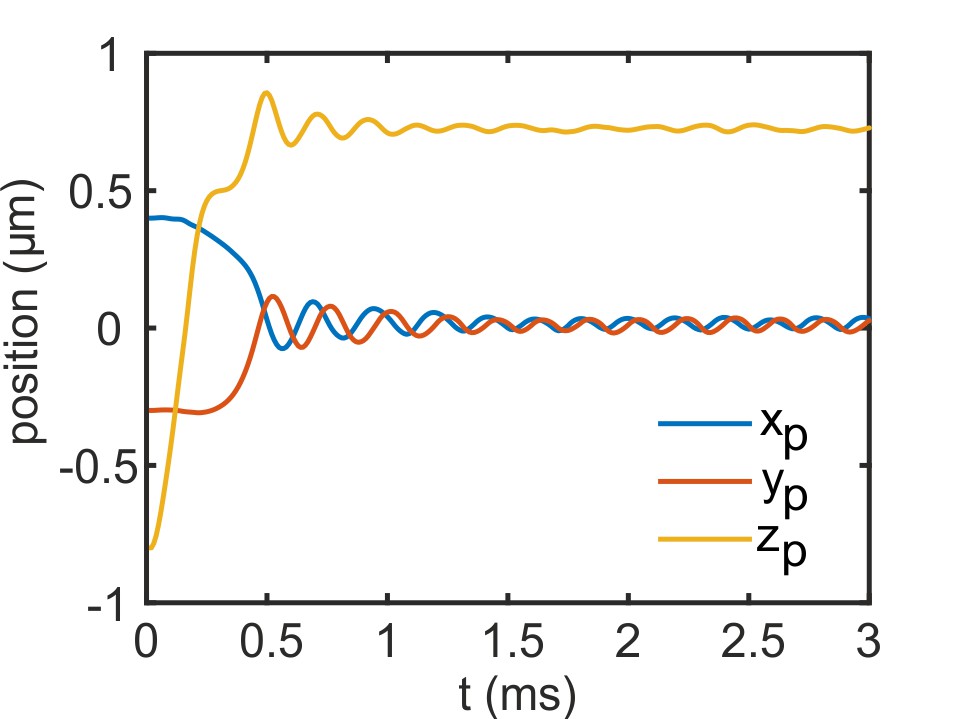}
           \caption[]{ 
           A NP is placed at an initial position ($400$ nm, $-300$ nm, $-800$ nm). We follow its trajectory during its trapping toward the final position ($0$ nm, $0$ nm, $725$ nm). The particle has a radius of $r_p=110$ nm and a mass of $m =1.05\times10^6 (4/3\pi r_p^3)$. A damping force was introduced of $-\gamma \dot{\mathbf{u}}$ where $\dot{\mathbf{u}}$ is the particle velocity, and $\gamma = 10^{-6}$ kg/s is the damping parameter. We assumed an incident wave with an intensity of $I_{0}=2.1\times10^{-4}\frac{W}{\mu\mathrm{m}^{2}}$, which means that for a Gaussian beam with a beam waist of  $W_0=5 \mu$m, we need a laser with an incident power of $P=16.4$ mW.}
     \label{fig:MotionPotential}
\end{figure}

As an example, we show in Fig.~\ref{fig:MotionPotential} the trapping trajectory of the NP for a given initial condition. In order to obtain a proper trapping volume, an incident field with a circular polarization was employed as  linearly polarized fields would yield an incomplete gradient. Furthermore, a reasonable estimate for the NP damping in air may be obtained in the range $10^{-6}-10^{-8}$~kg/s, as confirmed from experiments with a mechanical oscillator~\cite{jap2002}.
For most practical application of the presented concept in optical filtering, separation, and sorting, it is envisioned that large numbers of nanorings may need to be fabricated and integrated. Therefore, it is worthwhile to consider the influence of neighboring rings on the field properties of a given trapping site. For this reason, we investigated the focusing effect of a single ring but employed periodic boundary conditions in the lateral  directions ($xy$ plane). By monitoring the field magnitude at the location of the hot spot of an isolated ring as a function of the period, we obtained an estimate of a minimal ring separation distance of 800~nm  required to sufficiently suppress the coupling. Under such conditions,  each ring in the array acts as an individual trapping site, supporting the design principle depicted in Fig.~\ref{setup}.

In conclusion, the modelled concept of a dielectric ring based field modulation  theoretically proved to present a viable approach to optically trap dielectric NPs. The spatial modification of the electromagnetic field and its enhancement by a factor of 5, yielding a hot spot in the vicinity of the ring, was seen to provide a means to control the transport of nearby NPs. NPs distributed near the ring were shown to follow trajectories to entrapment under the engendered force field.
Considering the possibility of using metallic rings or other metallic nanostructures in conjunction with a dielectric ring and invoking thermoplasmonics~\cite{Passian:2006,Lereu:2008,Lereu:13}, one may envision applications where bacterial aerosolized particles may be  trapped and killed by being subjected to heat. Following the presented simulations, the special emphasis placed on the dielectric aspect of the traps can be assumed  crucial for avoiding excessive  absorption and thus heat generation which could impact biological applications.
\section{acknowledgements}
This research was supported in part by the laboratory directed research and development fund at Oak Ridge National Laboratory (ORNL). M.K. acknowledges support from Hector Fellow Academy. R.A. and C.R. acknowledge partial financial support by the Deutsche Forschungsgemeinschaft through CRC 1173.  ORNL is managed by UT-Battelle, LLC, for the U. S. DOE under Contract No. DE-AC05-00OR22725.



\end{document}